\documentstyle[prl,aps,multicol]{revtex}
\input{epsf}
\begin{document}
\draft
\title{Two interacting quasiparticles above the Fermi sea}

\author{Ph. Jacquod$^{(a)}$ and D. L. Shepelyansky$^{(b, *)}$}

\address {$^{(a)}$ Institut de Physique, Universit\'e de Neuch\^atel,
1, Rue A.L. Breguet, 2000 Neuch\^atel, Suisse \\
$^{(b)}$ Laboratoire de Physique Quantique, UMR C5626 du CNRS, 
Universit\'e Paul Sabatier, F-31062 Toulouse Cedex 4, France}

\date{\today}

\maketitle

\begin{abstract}
We study numerically the interaction and disorder effects for two 
quasiparticles in two and three dimensions. The dependence of the 
interaction-induced Breit-Wigner width $\Gamma$ on the excitation 
energy $\epsilon$ above the Fermi level, the disorder strength and 
the system size is determined. A regime 
is found where $\Gamma$ is practically independent of $\epsilon$.
The results allow to estimate the two quasiparticle mobility edge.
\end{abstract}
\pacs{PACS numbers: 72.15.Rn, 71.30+h}

% 72.15.Rn	Quantum localization
% 71.30+h	Metal-insulator transitions

\begin{multicols}{2}
\narrowtext

% Introduction

Recently the combined effect of interaction and disorder has been
studied by different groups for two particles in a random potential 
\cite{tip,imry,pich,oppen1,arcs}. These researches showed 
an interaction-induced enhancement of the two-particle localization 
length $l_c$ compared to the non-interacting length $l_1$. 
For low-dimensional systems (1d, 2d) the two interacting particles (TIP) can
propagate coherently over a large distance $l_c \gg l_1$, 
but still remain localized. In 3d,
the interaction can lead to TIP delocalization in an otherwise completely 
localized regime. In physical systems however 
one should study the interaction effect at finite
particle density. This type of problem is much more difficult for both the
analytical and numerical approaches. Up to now the only theoretical treatment
of this case has been done by Imry \cite{imry}. He took into account the effect
of the Fermi sea on the interaction-induced delocalization of two
quasiparticles. According to Imry's estimate in 3d the two quasiparticle 
mobility edge is lower compared to the non-interacting one. 
So far only the 1d case
has been treated numerically \cite{oppen2}. However, this case is of lesser
importance since the enhancement is recovered quite
far above the Fermi level contrarily to the 2d and 3d cases. Therefore it is
very important to study the problem in higher dimensions.

Exact numerical computations at finite density are quite heavy and due to that
only small system sizes are accessible \cite{didier,berko}. Even though this
approach has led to a number of interesting results, it seems that size
restrictions would not allow to check the Imry estimate since it requires a
relatively large one-particle localization length $l_1$. 
Therefore we chosed another
approach which is based on the computation of the interaction-induced
Breit-Wigner width $\Gamma$ of the local density of states of two interacting
quasiparticles (TIQ) above the Fermi sea. This width plays an
important role since it is directly related to the enhancement factor for the
localization length $\kappa \sim \Gamma \rho_c$, where $\rho_c$ is
the density of states coupled by the interaction \cite{tip,imry,gam}. Also
$\Gamma$ strongly affects the $\Sigma_2(E)$-statistics on energy scale $E > \Gamma$
\cite{pich2}.
Such $\Gamma$-based approach even though not direct is much more efficient 
and allowed to get a better understanding of TIP localization in the 1d 
case \cite{oleg}. To facilitate the numerical simulations we used the
approximation proposed in \cite{imry,oppen1,oppen2} based on the
consideration of only two quasiparticles above the Fermi energy $E_F$ 
neglecting all TIQ transitions involving hole excitations below $E_F$.
With such an approximation the quasiparticle lifetime becomes infinite or
in other words the inelastic processes are suppressed (see discussion below).
In this context we are able to study the TIQ problem in
blocks of linear size up to $L=30$ in 2d and $L=10$ in 3d which are significantly
larger than in exact diagonalization of multiparticle problems. Our approach
can also give a better understanding of the problem of quasiparticle 
interactions
in a quantum dot which has been recently addressed experimentally \cite{sivan}
and theoretically \cite{aronov}. 

For numerical studies we chose the TIQ model with on-site interaction of
strength $U$ on the 2d/3d Anderson lattice with intersite hopping $V$ and
diagonal disorder homogeneously distributed in the interval $ [-W,W]$.
The eigenvalue equation expressed in the basis of two-particle unperturbed
eigenstates reads 
%1
\begin{eqnarray}
(E_{m_1}+E_{m_2})\chi_{m_1, m_2} & + & 
U \sum_{{m^{'}_1}, {m^{'}_2}} Q_{m_1, m_2, {m^{'}_1}, {m^{'}_2}}
 \chi_{{m^{'}_1}, {m^{'}_2}} \nonumber \\
  & = & E\chi_{m_{1}, m_{2}}
\end{eqnarray}
%\indent
where $R$ determines the transformation between the lattice sites basis $| n
\rangle$ and the one-particle eigenbasis $\phi_m$ with eigenenergies $E_m$ so
that $|n \rangle = {\sum_{m} R_{n,m}} \phi_m$.
Accordingly $\chi_{m_{1}, m_{2}}$ are eigenfunctions of the TIQ problem
in one-particle eigenbasis. The matrix of transitions produced
by the interaction is 
$Q_{m_1, m_2, {m^{'}_1}, {m^{'}_2}} = 
{\sum_{n} R^{+}_{n, m_{1}} R^{+}_{n, m_{2}} R_{n, m^{'}_{1}} R_{n, 
m^{'}_{2}}}$. The Fermi sea is introduced by restricting the sum in (1) to
$m^{'}_{1,2}$ with unperturbed energies $E_{m^{'}_{1,2}} > E_F$. The
value of $E_F$ is determined by the filling factor $\mu$ which was 
fixed at $\mu = 1/4$ in 2d and $\mu = 1/3$ in 3d. However the results are
not sensitive to this choice. At small disorder this gives
approximately $E_F \approx -1.4V$ and $E_F \approx -V$ respectively. Due to the
on-site nature of the interaction, only symmetric configurations were
considered. By direct diagonalization of the model (1) we computed the local
density of states
\begin{equation}
\rho_{W} (E-E_{m_1}-E_{m_2}) = 
\sum_{\lambda} \mid \chi^{(\lambda)}_{m_1, m_2} \mid^2 \delta(E-E_\lambda)
\end{equation}
This function characterizes the probability contribution of eigenfunction
$ \chi^{(\lambda)}_{m_1, m_2} $ with eigenenergy $E_\lambda$ in the
unperturbed basis $|\phi_{m_1} \phi_{m_2} \rangle$. Generally we found that
it has the well-known Breit-Wigner shape $\rho_{BW} (E) = \Gamma/(2 \pi
(E^2+\Gamma^2/4))$ (see Fig.1) where the width $\Gamma$ depends on 
the parameters of
the model. Our main aim was to investigate this dependence on the system size,
the interaction strength and the TIQ excitation energy
$\epsilon=E-2E_F$ above the Fermi sea.

\begin{figure}
\epsfxsize=3in
\epsfysize=2in
\epsffile{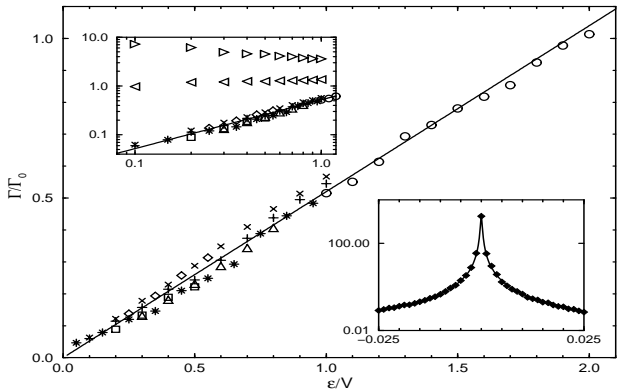}
\vglue 0.2cm
%\medskip
\caption{Energy dependence of the rescaled Breit-Wigner width 
$\Gamma/\Gamma_0$ in 2d. Direct diagonalization (DD)  data  at $W/V=1$:
$U/V=0.6$ with $L=15$ (triangles), $L=20$
(squares); $U/V=1.5$ and $L=20$ (diamonds). Fermi golden rule (FGR) data:  
$W/V=1$ with $L=20$ ($+$), $L=25$ ($\times$); $W/V=0.5$ with $L=15$ (*).
The straight line $\Gamma(\epsilon)/\Gamma_0 = C
\epsilon/V$ with $C=0.52$ shows the Imry estimate.
Upper insert: the same in a log-log scale with FGR data at 
higher disorder $W/V=3$ (left) and $W/V=5$ (right) triangles ($L=30$).
Lower insert: $\rho_W$ vs. $E$ for $L=20, W=V=1, U=0.6, \epsilon =0.4$
fitted by $\rho_{BW}$ with $\Gamma=0.18 \Gamma_0$ (solid curve).
} 
\label{fig1}
\end{figure}

The results for the 2d, 3d cases in the regime of weak disorder are presented
in Figs.1, 2. For sufficiently high excitation 
energy $\epsilon$, the restriction
imposed by the Fermi sea becomes unimportant and the width $\Gamma(\epsilon)$
tends to the value $\Gamma_0 = U^2/(VL^d)$ which corresponds to the result
obtained with ergodic eigenfunctions \cite{tip,imry,gam}. In this approach the
transition matrix elements have a typical value 
$U_s^2 = U^2 Q^2 \sim U^2/L^{3d}$ and the
transition rate is given by the Fermi golden rule with $\Gamma_0 \sim U^2 Q^2
\rho_c$ and $\rho_c \sim L^{2d}/V$. The presence of 
the Fermi sea modifies this density
which becomes energy-dependent $\rho_c(\epsilon) \sim L^{2d} \epsilon/V^2$. 
As a result the width $\Gamma$ drops with decreasing energy as \cite{imry}
\begin{equation}
\Gamma(\epsilon) = C \Gamma_0 \frac{\epsilon}{V} = 
C \frac{U^2 \epsilon}{V^2 L^d}
\end{equation}
This behaviour was assumed to remain valid
for weak enough disorder as long as $L \leq l_1$. 
Hence $\Gamma$ is independent on the disorder strength 
$W$. Indeed for $l_1 \gg L$ this estimate is in good agreement with the 
numerical data presented in Figs.1, 2 with
$C=0.52$ (2d) and $C=0.3$ (3d). Most of the data for 
$\Gamma$ in Figs.1, 2 were obtained by 
direct diagonalization of the model (1). Another way 
to determine $\Gamma$ without computation of the TIQ eigenstates is based on
the Fermi golden rule which should remain valid for moderate interaction
strength. This approach gives $\Gamma = 2 \pi \sum_{m_1',m_2'}
\mid U Q_{m_1,m_2;m_1',m_2'} \mid^2 \delta(\epsilon+2E_F-E_{m_1'}-E_{m_2'})$
in terms of the transition matrix elements between one-particle eigenstates only.
Here $\epsilon = E_{m_1}+E_{m_2}-2E_F$ and to improve the statistics we
averaged over different $m_{1,2}$ with approximately the same $\epsilon$.
As can be seen in Figs.1, 2, both methods are in good agreement for
interaction strength $U \leq 1.5 V$. Another confirmation of the validity of
the golden rule is the $U^2$-dependence of $\Gamma$ obtained by direct
diagonalization (Fig.2). Both approaches confirm also the scaling 
$\Gamma \propto L^{-d}$ valid for weak disorder. We used up to 100
realizations of disorder for the Fermi golden rule approach and up to 500
for direct diagonalization.

\begin{figure}
\epsfxsize=3in
\epsfysize=2in
\epsffile{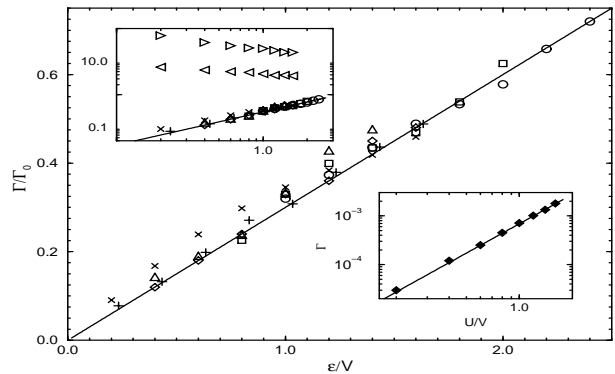}
\vglue 0.2cm
%\medskip
\caption{Same as Fig.1 in 3d. DD data at $U/V=1.2$ and $W/V=2$ with
$L=4$ (o), $L=5$ (squares), $L=6$ (diamonds) and $L=7$ (triangles).
FGR data: $W/V=2$ with $L=10$ ($\times$);
$W/V=1$ with $L=8$ ($+$). Here $C=0.3$. Upper insert: the same in a log-log
scale with FGR data at higher disorder $W/V=6$ (left) and $W/V=10$ (right)
triangles (L=10). Lower insert: FGR data for $\Gamma$ vs. $U/V$
at $W/V=2, L=6, V=\epsilon=0.5$
(solid line: $\Gamma=0.3 \Gamma_0$).
} 
\label{fig2}
\end{figure}

The situation becomes more intricate at higher disorder. Here our results show
that $\Gamma$ becomes much less sensitive to $\epsilon$ variation
(Figs.3). There is a clear tendency that at still moderate
disorder $\Gamma$ becomes practically independent on $\epsilon$
which has been varied over one order of magnitude. In the 3d
case, such behaviour takes place even in the delocalized regime $W < W_c
\approx 8.2 V$. The data
even indicate a small growth of $\Gamma(\epsilon)$ with decreasing $\epsilon$
at $W \geq 6 V$. At high disorder $l_1$ decreases
and becomes comparable or even less than $L$. In this
situation the eigenstates are no longer ergodic in the block and the scaling
$\Gamma \propto L^{-d}$ is no more valid. In the limit $1 < l_1 \ll L$, it is
natual to expect another scaling $\Gamma \propto l_1^{-d}$. To check this
scaling we computed the inverse participation ratio (IPR) $\xi \sim l_1^d$
which allowed to calculate the ergodic value $\Gamma_{1} = U^2/(V \xi)$.

\begin{figure}
\epsfxsize=3in
\epsfysize=2in
\epsffile{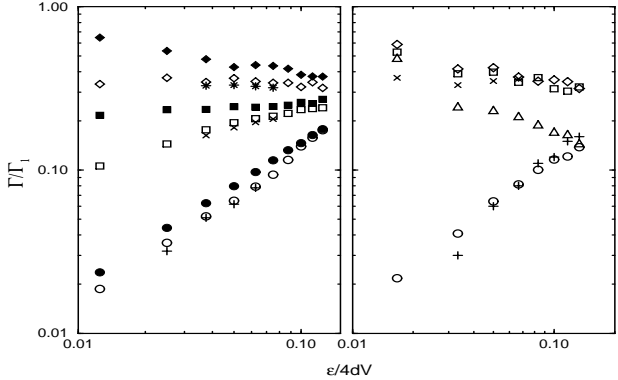}
\vglue 0.2cm
%\medskip
%\hspace{0.2cm}
\caption{
Energy dependence of the rescaled Breit-Wigner width $\Gamma/
\Gamma_1$ in 2d (left) and 3d (right). FGR data in 2d: $W/V=1$ with
$L=20$ (o) and $L=30$ (points); $W/V=2$ with $L=20$ (open squares) and
$L=30$ (full squares); $W/V=3$ with $L=20$ (open diamonds)
and $L=30$ (full diamonds). DD data in 2d at $L=20$, $U/V=0.6$: $W/V=1$ (+), 
$W/V=2$ ($\times$) and $W/V=3$ (*).  FGR data in 3d at $L=10$: $W/V=2$ (o),
$W/V=6$ (squares), $W/V=10$ (diamonds) and $W/V=14$ (triangles).
DD data in 3d at $L=6$, $U/V=1.2$: $W/V=2$ (+) and $W/V=6$ ($\times$).
} 
\label{fig3}
\end{figure}

At sufficiently high excitation energy, the real
width should be $\Gamma \sim \Gamma_{1}$ which
gives the correct scaling with system size in the localized regime according
to the numerical data. This would
explain why in the block of fixed size $\Gamma$ increases with increasing
disorder (see inserts in Figs.1, 2). While such estimate gives the correct
value of $\Gamma$ at high energies $\epsilon \approx 2 V$ (Fig.3), 
it however does not explain the change of energy dependence with disorder.
We should note that even in this unusual regime both the direct diagonalization 
and the Fermi golden rule
computations give the same values of the width $\Gamma$.

It is clear that the change of the energy dependence of $\Gamma$ cannot be
explained in the framework of ergodic transition matrix elements $U_s^2 \sim
(U/V)^2 \Delta^3/V$ where $\Delta \sim V/L^d$ is the one-particle level spacing
in the block of size $L$. At the same time the numerical 
results for the two-particle
density of states $\rho_2$ definitely show that it increases linearly with the
excitation energy $\epsilon$ as $\rho_2 \sim \epsilon/\Delta^2$. Therefore
the only possibility is that at higher disorder the ergodic estimate for $U_s$
is no longer valid. Indeed from the theory of quasiparticle lifetime in
disordered metals and quantum dots \cite{shmid,altshuler,aronov} it is known
that the diffusive nature of the dynamics should be taken into account. For 
excitation energy $\epsilon \gg E_c$ much bigger than the Thouless energy
$E_c$, the quasiparticle decay rate is $\Gamma_D \sim U_s^2 \rho_3 \sim \Delta
(U \epsilon/V E_c)^{d/2}$ where 
$\rho_3 \sim \rho_2 \epsilon/\Delta$ is the density
of three-particle states composed of two particles and one hole in the final
state. In the other regime relevant for the metallic quantum dot 
$\Delta < \epsilon \ll E_c$ this rate is $\Gamma_D \sim \Delta
(U \epsilon/V E_c)^2$ \cite{aronov}. 
This shows that the matrix elements $U_s^2
\sim \Gamma_D/\rho_3$ are not always given by the ergodic estimate in agreement
with recent results \cite{blan}. The different nonergodic regimes
can be described by the following approximate expression \cite{kamenev}
\begin{equation}
U_s^2 \sim \left( \frac{U}{V} \right)^2 \frac{\Delta^2}{g^2} \left(
1+\frac{\epsilon}{E_c} \right)^{d/2-2}
\end{equation}
where $g=E_c/\Delta$ is the conductance assumed to be much bigger
than one. According to (4) the TIQ width $\Gamma \sim U_s^2 \rho_2$
increases with disorder $W$ even in the metallic regime since $E_c = D/L^2 \sim
V^3/(W L)^2$ with $D$ being the diffusion constant. The ergodic estimate for
$U_s^2$ is recovered for $g > E_F/\Delta \sim V/\Delta$ \cite{blan}
corresponding to very weak disorder. While the exact numerical coefficients
in (4) are not known, it gives the energy-dependence $\Gamma \propto
\epsilon^{d/2-1}$ which is in agreement with the numerical data for 
$d=2$ (Fig.3) but in 3d the data indicate an
algebraic dependence with power $\alpha < 0$ ($\alpha \approx -0.2$ for $W=6V$
and $\alpha \approx -0.3$ for $W=10V$) instead of the theoretical value $\alpha
= 1/2$. There can be different reasons for this discrepancy. One case $W=10V$ 
corresponds to the localized regime while the theory requires a metallic
behaviour. The other case $W=6V$, even though still delocalized, is quite
close to the critical value $W_c$. Our data indicate that in the metallic
regime with $1 < W/V < 6$ the power $\alpha$ smoothly changes from 1 to -0.3.

Surprisingly at present there are no theoretical
predictions for $U_s^2$ not only near the critical value $W_c$ but also in
the localized regime with $l_1 \gg 1$. It seems natural to make the assumption
that in the localized case, the transition matrix elements will be given
by an equation similar to (4) with $g \approx 1$ since in a block of size $l_1$
the Thouless energy is $E_c \approx \Delta \sim V/l_1^d$. This gives $\Gamma
\sim \Gamma_1 (\epsilon/\Delta)^\alpha$ where $\alpha$ has replaced the
theoretical value $d/2-1$ valid in the metallic regime. 
We will assume that for $d \geq 2$ the exponent 
$\mid \alpha
\mid < 1$. In 3d the TIQ mobility edge $\epsilon_{m2}$ is defined by the 
condition $\kappa=\Gamma \rho_c > 1$ \cite{tip,imry}. 
Since $\rho_c \sim \epsilon / \Delta^2$ the above
expressions for $\Gamma$ give 
\begin{equation}
\epsilon_{m2} \sim \frac{V}{l_1^d} \left( \frac{V}{U} \right)^{2/(1+\alpha)}
\sim V \left( \frac{\epsilon_{m1}}{V} \right)^{\nu d} 
\left( \frac{V}{U} \right)^{2/(1+\alpha)}
\end{equation}
where $\epsilon_{m1} \sim V l_1^{-1/\nu}$ is the one-particle mobility edge. 
The one-particle critical exponent is $\nu \approx 1.5$. 
Due to that for $U \sim V$, the edge
$\epsilon_{m2} \ll \epsilon_{m1}$.  The above result (5) gives for
$\epsilon_{m2}$ a much smaller value than the one given by the Imry estimate
\cite{imry}. The main reason for this is that the transition matrix elements
in the block of size $l_1$ where $g \approx 1$ are much larger than their
ergodic value used in \cite{imry}. The condition that the
TIQ delocalization border in $U$ at $\epsilon \sim V$ is the same as
for TIP ($U>V/{l_1}^{d/2}$ \cite{tip,imry}) gives $\alpha=0$.

The numerical results for the dependence of $\kappa = \Gamma \rho_c$ on 
$\epsilon$ are
presented in Fig.5. To determine numerically the density of coupled states
$\rho_c \sim \epsilon l_1^{2d}/V^2$ in the localized regime we computed it 
taking into account only those TIQ states which give contributions 
larger than 30\% of the value of $\Gamma$ (the data were not sensitive
to the cut-off value). The density $\rho_c$ 
defined in this way
is independent of the system size when $L > l_1$. The data show that
for $W=14V$ there is no TIQ delocalization 
($\kappa < 1$). However closer to the one-particle delocalization border but
still above it ($W > W_c$) the value of $\kappa$ becomes bigger than one
at moderate excitation energies and TIQ delocalization should take place.
Further investigations are required to check more accurately the theoretical
prediction (5) for the mobility edge $\epsilon_{m2}$.

\begin{figure}
\epsfxsize=3in
\epsfysize=2in
\epsffile{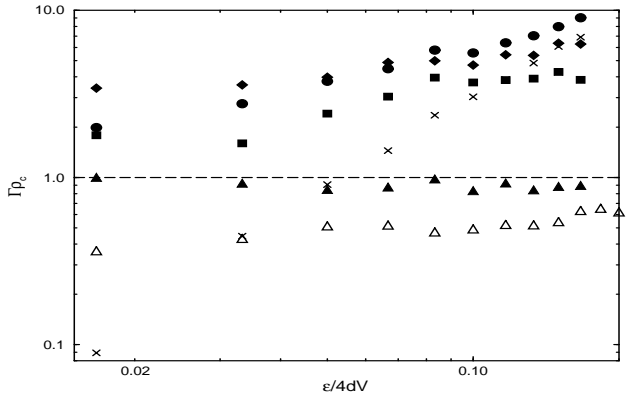}
\vglue 0.2cm
%\medskip
\caption{
Energy dependence of the TIQ delocalization parameter 
$\kappa = \Gamma \rho_c$ in 3d (FGR data) at $U/V=1.2$:
$L=7$ for $W/V=14$ (open trianges);
$L=10$ for $W/V=2$ ($\times$), $W/V=6$ (points),
$W/V=8.4$ (squares), $W/V=10$ (diamonds) and $W/V=14$ (full triagles).
The dashed line shows the TIQ mobility edge $\kappa = \Gamma \rho_c =1$.
} 
\label{fig5}
\end{figure}

In our model (1) all inelastic processes involving hole excitations have been
suppressed so that $\Gamma_D=0$. In a general case when no special physical 
effects suppress $\Gamma_D$ the ratio $\Gamma_D/\Gamma \sim \rho_3/\rho_2 \sim
\epsilon/\Delta$ is larger than one for $\epsilon \gg \Delta$. This means that
the quasiparticle lifetime is shorter than the typical time of TIQ transitions.
However according to recent predictions \cite{gefen} the value of $\Gamma_D$ in
a quantum dot is zero for excitation energies $\epsilon < \epsilon_1 \sim
\Delta \sqrt{g}$ for $U \sim V$. This result was derived for metallic quantum 
dots with $g \gg
1$. It is not clear what will be the result for dots with $g \sim 1$ (the
blocks of size $l_1$ should be considered for TIQ delocalization \cite{imry})
but it is natural to assume that $\epsilon_1 \sim \Delta \sim
V/l_1^d$. This border is 
parametrically comparable to the TIQ mobility edge $\epsilon_{m2}$ (5). The
numerical value of the ratio $\epsilon_{m2}/\epsilon_1$ is not known. It is
possible that it is less than one due to the 1/3-drop of energy per particle
in the one-particle decay process. In this situation the quasiparticle
interaction can be more important than the one-particle decay.

In conclusion we analyzed the interaction-induced delocalization of two
quasiparticles above the Fermi sea in the approximation where inelastic
processes are suppressed. According to our numerical results and
semi-analytical estimates, the TIQ mobility edge is below the one-particle
edge ($\epsilon_{m2} < \epsilon_{m1}$). The obtained value of $\epsilon_{m2}$
is even much smaller than the value proposed in \cite{imry} due
to the fact that transition matrix elements are much bigger than their ergodic
value near the Fermi level.

We thank O. P. Sushkov for useful discussions. 
One of us (P.J.) 
gratefully acknowledges the hospitality of the Laboratoire de Physique
Quantique, Universit\'e Paul Sabatier.

\end{multicols}

\end{document}